\documentclass[12pt]{iopart}
\usepackage{epsfig}
\begin{document}
\title[Real and Virtual Photon Propagation]{Radiating and Non-Radiating
Current Distributions in Quantum Electrodynamics}
\author{G. Castellani, S. Sivasubramanian,
A. Widom\footnote[3]{To whom correspondence should
be addressed (widom@neu.edu)}, Y.N. Srivastava}
\address{Physics Department, Northeastern University, Boston,
MA 02115, USA}

\begin{abstract}
The notion of ``radiating'' and ``non-radiating'' current sources in
classical electrodynamics plays an important role in calculations of
direct and inverse electromagnetic scattering problems. Such a
decomposition of the current is central for the notion of localized
non-radiating electromagnetic modes. A completely quantum electrodynamic
view is explored in this work. Photon emission and absorption current
sources are classified as being either radiating or non-radiating.
This quantum classification corresponds, respectively and exactly,
to the notion of ``real'' and ``virtual'' photon processes.
Causal properties of both real and virtual electromagnetic fields are
discussed.
\end{abstract}
\maketitle

\section{Introduction}

In both classical electrodynamics and quantum electrodynamics, an
electric current source \begin{math} J^\mu \end{math} produces the
electromagnetic field \begin{math} F_{\mu \nu} \end{math} via
Maxwell's equations
\begin{equation}
\partial_\mu \ ^*F^{\mu \nu}=0,
\ \ \ \partial_\mu F^{\mu \nu}=-(4\pi /c)J^\mu .
\end{equation}
The nature of the microscopic current \begin{math} J^\mu \end{math}
has always been of central physical interest, e.g. in the
context of direct and inverse electromagnetic
scattering\cite{Bleistein_77,Vigoureux_92,Carney_00,Schotland_01}.

Some classical current distributions radiate energy while others do not.
In early works\cite{Sommerfeld_05,Herglotz_08,Hertz_08}, it was established
that self oscillations existed without radiation for some extended
electron models. For example, a single point charge moving uniformly in a
circle with period \begin{math} T \end{math} will radiate electromagnetic
energy. If the point charge were replaced by a spherical shell
\cite{Schott_33,Schott_37} of diameter \begin{math} d \end{math},
then no radiation will be present if \begin{math} d=ncT \end{math} where
\begin{math} n=1,2,3,\ldots \ \end{math} and if the orbital motion diameter is
less than \begin{math} d \end{math}. Spherical shell charge distributions
may thus wobble indefinitely without damping forces, radiative losses
or any external forces\cite{Markov_46,Bohm_48}.

The above kind of non-radiating charge distribution
acts similarly to an ideal radiation cavity\cite{Goedecke_68} which does not
leak the radiation field to infinity. The classification of classical
currents into radiating and non-radiating parts has been thoroughly discussed
in the literature\cite{Wolf_73,Friedlander_73, Hoenders_79,Kim_86,Gamliel_89,Meyer_89,Hoenders_98,Marengo_99,Marengo_00}.
Classical ``point'' charges in accelerated motion will always produce some
radiation\cite{Schott_33}. Moving and/or spinning radially symmetric charge
distributions as well as radially asymmetric charges distributions
can be non-radiating\cite{Ksienski_61,Goedecke_64,Erber_70}. Smooth current
distributions may or may not radiate. Note that quantum mechanical
current distributions are always smooth. The very early work on non-radiating
classical current sources\cite{Sommerfeld_05,Herglotz_08,Hertz_08} was carried
out in the pre-quantum mechanical context of understanding why electrons
did not fall into nucleus. This problem is presently understood on
the basis of the stability of the atomic or molecular quantum
mechanical ground states.

For the quantum electrodynamic case of atoms, molecules or nuclei, charged
currents may some times (but not always) radiate electromagnetic energy,
e.g. during an energy decay process. In other quantum mechanical cases,
the currents produce electromagnetic fields only in the neighborhood of
the source but without emitting radiation into a ``far zone''. The
magnetic field of an atom with non-zero total angular
momentum\cite{Lieb_86a,Lieb_86b}
\begin{math} \hbar\sqrt{J(J+1)} \end{math} in the ground state
serves as an example of a non-radiating current source described by a
magnetic moment.

Our purpose is to show that the well developed notion of
radiating and non-radiating currents in classical electrodynamics
corresponds exactly to the notion of ``real'' and ``virtual'' photons
in quantum electrodynamics. The charged currents will be classified
in terms of radiating parts which produce
{\em real} photons and non-radiating parts which produce {\em virtual}
photons. In Sec.2, the notion that classical currents emit photons
with a Poisson distribution will be reviewed\cite{Sudarshan_68}.
The mean number of photons will be computed in terms of the photon
propagator. The {\em radiating} part of the current source will be
classified as that part which emits {\em real} photons.
In Sec.3 it will be shown that the non-radiating parts of
the currents correspond to {\em virtual} photon exchange which
can control Coulomb and Ampere forces but which do not actually radiate.
In Sec.4, the general classification of current matrix elements
emitting {\em real} and {\em virtual} photons will be performed.
In Sec.5, causality will be discussed. In the concluding Sec.6,
the physical properties of virtual electromagnetic fields and real
electromagnetic fields will be explored from a quantum electrodynamic
viewpoint. Virtual photons, which yield advanced and
retarded (at a distance) current interactions, travel on the light
cone\cite{Wheeler_45}. Propagation on the light cone in space-time
corresponds to propagation off the energy shell in
energy-momentum variables. {\em Real} photons travel strictly on the
energy shell and thereby propagate off the space-time light cone.

\section{Emission of Real Photons}

Classical current sources when coupled into the quantum
electrodynamic fields emit a mean number
\begin{math} \bar{N} \end{math} of photons. Assume that the quantum
state of the photons is the vacuum \begin{math} \left|0\right> \end{math}
{\em before} the classical current sources begin to radiate.
The probability of having emitted
\begin{math} N \end{math} photons after the current sources are
finished radiating is given by the Poisson distribution
\begin{equation}
P_N=\frac{\bar{N}^N e^{-\bar{N}}}{N!}\ .
\label{PD1}
\end{equation}
In particular, the probability that the vacuum (before a
classical current pulse) remains the vacuum (after a current pulse)
is given by
\begin{equation}
P_0=e^{-\bar{N}}\ .
\label{PD2}
\end{equation}
The vacuum action functional \begin{math} S[J] \end{math} of a
classical current source is defined by the amplitude that the vacuum
(before a classical current pulse) remains the vacuum
(after a current pulse); i.e.
\begin{equation}
\exp \left(\frac{iS[J]}{\hbar }\right)=
\left<0\left|\left\{\exp
\left(\frac{i}{\hbar c^2}\int J^\mu (x) \hat{A}_\mu (x)d^4x\right)
\right\}_+\right|0\right>,
\label{PD3}
\end{equation}
wherein \begin{math} + \end{math} denotes time ordering,
\begin{math} \hat{A}_\mu (x) \end{math} denotes the operator vector
potential and \begin{math} J^\mu (x) \end{math} describes the classical
current source. The vacuum to vacuum probability is the absolute value
squared of the vacuum to vacuum amplitude
\begin{equation}
P_0=\left|e^{iS[J]/\hbar } \right|^2=e^{-2{\Im m}S[J]/\hbar }.
\label{PD4}
\end{equation}
From Eqs.(\ref{PD2}) and (\ref{PD4}) it follows that the imaginary
part of the action determines the mean number of photons emitted
by classical current sources; i.e.
\begin{equation}
\bar{N}=\frac{2}{\hbar }{\Im m}S[J].
\label{PD5}
\end{equation}

Since photon quantum oscillators in the ground state yield Gaussian
distributions for the vector potential, the action as determined by
the photon propagator
\begin{equation}
D_{\mu \nu }(x,y)=\frac{i}{\hbar c}
\left<0\left|\left\{\hat{A}_\mu (x)\hat{A}_\nu (y)\right\}_+\right|0\right>
\label{PD6}
\end{equation}
is
\begin{equation}
S[J]=\frac{1}{2c^3}\int \int
J^\mu (x)D_{\mu \nu }(x,y)J^\nu (y)d^4x d^4y.
\label{PD7}
\end{equation}
The photon propagator depends on the choice of the vector potential
gauge. In the Feynman gauge,
\begin{equation}
D_{\mu \nu }(x,y)=\int
\frac{4\pi \eta_{\mu \nu} }{Q^2-i0^+}
\ e^{iQ\cdot (x-y)}\ \frac{d^4Q}{(2\pi )^4}\ .
\label{PD8}
\end{equation}
It is shown in the Appendix A that
\begin{equation}
D_{\mu \nu }(x,y)=
\frac{i\eta_{\mu \nu }}{\pi \{(x-y)^2+i0^+\}}\ .
\label{PD8a}
\end{equation}

Defining the Fourier transform of the current sources as
\begin{equation}
J^\mu_Q=\int e^{-iQ\cdot x} J^\mu (x) d^4x
=(J^{\mu}_{-Q})^* ,
\label{PD9}
\end{equation}
Eqs.(\ref{PD5}) and (\ref{PD7})-(\ref{PD9}) imply
the central result of this section; i.e. the mean number of photons
radiated by a classical current source is given by
\begin{equation}
\bar{N}=\frac{1}{4\pi^2 \hbar c^3}
\int \left(\eta_{\mu \nu }J^\mu_{-Q}J^\nu_Q \right)
\delta (Q^2) d^4Q.
\label{PD10}
\end{equation}
From Eq.(\ref{PD10}) we have the following:
\par \noindent
{\bf Theorem 1:} {\em Real photons are radiated (i) only by that
part of the current whose wave vector
\begin{math} Q=({\bf k},\omega /c) \end{math}
is on the ``zero mass shell''
\begin{equation}
Q^2=|{\bf k}|^2-(\omega /c)^2=0,
\label{PD11}
\end{equation}
and (ii) only by the spatial transverse part of vector
current; i.e.
\begin{equation}
{\bf J}^{trans}_{Q}=\left({\bf 1}-\frac{\bf kk}{|{\bf k}|^2}\right)
\cdot{\bf J}_{Q}=
\frac{{\bf k}\times ({\bf J}_Q\times {\bf k})}{|{\bf k}|^2}\ .
\label{PD12}
\end{equation}}
\par \noindent
{\bf Proof:} (i) Eq.(\ref{PD11}) follows directly from the
\begin{math} \delta (Q^2) \end{math} in the integral on the
right hand side of Eq.(\ref{PD10}). (ii) Conservation of charge
\begin{math} \partial_\mu J^\mu =0 \end{math} in space-time may
be Fourier transformed into
\begin{equation}
Q_\mu J^\mu _Q = {\bf k}\cdot {\bf J}_Q-\omega \rho_Q=0.
\label{PD13}
\end{equation}
Thus,
\begin{eqnarray}
\eta_{\mu \nu }J^\mu_{-Q}J^\nu_Q &=&
{\bf J}_{-Q} \cdot {\bf J}_Q-c^2\rho_{-Q} \rho_Q ,
\nonumber \\
&=& {\bf J}_{-Q} \cdot {\bf J}_Q-
(c/\omega )^2{\bf k\cdot J}_{-Q}{\bf k\cdot J}_Q ,
\nonumber \\
&=&{\bf J}_{-Q} \cdot {\bf J}_Q-
\frac{{\bf k\cdot J}_{-Q}{\bf k\cdot J}_Q}{|{\bf k}|^2}\ ,
\label{PD14}
\end{eqnarray}
where Eq.(\ref{PD11}) has been invoked.
From Eqs.(\ref{PD12}) and (\ref{PD14}), it follows
that
\begin{equation}
\left(\eta_{\mu \nu }J^\mu_{-Q}J^\nu_Q \right)\delta (Q^2)=
\left({\bf J}^{trans}_{-Q} \cdot {\bf J}^{trans}_Q\right)\delta (Q^2),
\label{PD15}
\end{equation}
hence the theorem is proved. This theorem completely characterizes
the part of a classical current distribution which contributes
to the emission of a mean number \begin{math} \bar{N} \end{math}
of real photons.

The quantum electrodynamic notion of real photon emission
here implies exactly the same conditions previously
employed\cite{Wolf_73} to characterize radiating classical current
sources. While we shall in what follows also consider the more
general case of quantum current sources, the recovery of the
known classical results by identifying real photons with classical
radiating currents is quite satisfactory. The quantum electrodynamic
notion of {\em virtual} photons will be identified in what follows
with the non-radiating part of the current sources. To see this in
detail we must consider the real part of the action in
Eq.(\ref{PD7}).

\section{Exchange of Virtual Photons}

While the imaginary part of the action determines the mean number of
real photons emitted via Eq.(\ref{PD5}), the real part of the action,
\begin{equation}
W[J]={\Re e}S[J]=
\frac{1}{2c^3}\int \int
J^\mu (x)\{{\Re e}D_{\mu \nu }(x,y)\}J^\nu (y)d^4x d^4y.
\label{VP1a}
\end{equation}
determines the interaction between currents due to photon exchange.
In detail, the action implicit in Eqs.(\ref{PD8a}) and (\ref{VP1a})
\begin{equation}
W[J]=
\frac{1}{2c^3}\int \int \delta \big((x-y)^2\big)
J^\mu (x)J_\mu (y)d^4x d^4y.
\label{VP1b}
\end{equation}
describe the motion of {\em virtual} photons
{\em on the light cone} between regions of space time
in which there are non-radiating
(\begin{math} Q^2\ne 0 \end{math})
parts of the current sources.

The function \begin{math} \delta \big((x-y)^2\big) \end{math}
on the right hand side of Eq.(\ref{VP1b}) (which
restricts the {\em virtual} photon to move on the light cone)
is due to the off zero mass shell part of the propagator. To see
what is involved, note that Eqs.(\ref{PD9}), (\ref{VP1a}) and
(\ref{A9}) imply that
\begin{equation}
W[J]=\frac{1}{(2\pi c)^3}\int
\left(\frac{\eta_{\mu \nu}J^\mu _{-Q}J^\nu _Q}{Q^2}\right)d^4Q
\ \ {\rm with}\ \ Q^2\ne 0.
\label{VP2}
\end{equation}
Eq.(\ref{VP1b}) may be written more explicitly as
\begin{eqnarray}
T &=& (t_1-t_2) \ \ {\rm and}\ \
{\bf R}=({\bf r}_1-{\bf r}_2), \nonumber \\
t_1 &=& t+\frac{T}{2} \ \ {\rm and}\ \
t_2=t-\frac{T}{2}. \nonumber \\
W[J] &=& -\int \int \int \bar{v}({\bf r}_1,{\bf r}_2,t)
d^3{\bf r}_1d^3{\bf r}_2dt = -\int U dt,
\label{VP3}
\end{eqnarray}
with an ``action at a distance'' potential
\begin{equation}
\bar{v} =\frac{c}{2}\int \delta (c^2T^2-R^2)
\left(\rho ({\bf r}_1,t_1)\rho ({\bf r}_2,t_2)
-\frac{{\bf J}({\bf r}_1,t_1)\cdot {\bf J}({\bf r}_2,t_2)}
{c^2}\right)dT.
\label{VP4}
\end{equation}
Formally, Eq.(\ref{VP4}) describes the interaction within charge
and current distributions in terms of the classical (one half)
retarded and (one half) the advanced propagators; i.e.
\begin{equation}
\delta (c^2T^2-R^2)=\frac{1}{2R}
\left\{\delta (cT-R)+\delta(cT+R)\right\}.
\label{VP5}
\end{equation}
For the special case of electro- and magneto- statics,
\begin{eqnarray}
\rho ({\bf r},t) &=& \rho ({\bf r})\ \ ({\rm only}),
\nonumber \\
{\bf J} ({\bf r},t) &=& {\bf J} ({\bf r})\ \ ({\rm only}),
\nonumber \\
div{\bf J}({\bf r}) &=& 0\ \ \ ({\rm statics}),
\label{VP6}
\end{eqnarray}
the total energy
\begin{math}
U=\int \int
\bar{v}({\bf r}_1,{\bf r}_2)d^3{\bf r}_1d^3{\bf r}_2
\end{math}
follows from Eqs.(\ref{VP4}), (\ref{VP5}) and (\ref{VP6}) to be
\begin{equation}
U=\frac{1}{2}\int \int \left\{
\frac{\rho ({\bf r}_1)\rho ({\bf r}_2)-
{\bf J} ({\bf r}_1)\cdot {\bf J} ({\bf r}_2)/c^2}
{|{\bf r}_1-{\bf r}_2 |}\right\}
d^3{\bf r}_1d^3{\bf r}_2 \ \ ({\rm statics}).
\label{VP7}
\end{equation}
The conventional photon exchange derivation of the Coulomb
plus Ampere energies, respectively, from charge and current
interactions has been reviewed and exhibited in Eq.(\ref{VP7}).

The exchanged photon interaction is
(i) ``on the light cone'' as in Eq.(\ref{VP4}) and
(ii) ``off the energy shell'' as in
Eq.(\ref{VP2}). These are the characteristics of a
{\em virtual} electromagnetic field. Furthermore, it is only the
virtual photons that move on the light-cone with velocity
\begin{math} c \end{math}.

\section{Radiating Current Matrix Elements}

Let us now consider in a fully quantum electrodynamics fashion that
part of the current {\em operator} which radiates or absorbs photons.
For example with \begin{math} \gamma  \end{math} as a finally radiated
photon in the reaction
\begin{equation}
I\to F+\gamma
\label{RJ1}
\end{equation}
one may consider exact "in" and "out" eigenstates of the four momenta
\begin{eqnarray}
P^\mu \left|I(in)\right> &=& P^\mu_I \left|I(in)\right>,
\nonumber \\
P^\mu \left|F(out)\right> &=& P^\mu_F \left|F(out)\right>.
\label{RJ2}
\end{eqnarray}
The current matrix element describing a real photon decay in
Eq.(\ref{RJ1}) is defined as
\begin{equation}
J^\mu_{FI}(x)=\left<F(out)\right|J^\mu (x)\left|I(in)\right>.
\label{RJ3}
\end{equation}
Since Eq.(\ref{RJ1}) describes real photon emission, (i) four momentum
is strictly conserved and (ii) the photon four momentum is
strictly on the energy shell
\begin{equation}
\hbar Q=P_I-P_F\ \ \Rightarrow\ \ \hbar^2 Q^2=(P_I-P_F)^2=0.
\label{RJ4}
\end{equation}
Since the Heisenberg current operator obeys
\begin{equation}
J^\mu (x)=e^{-iP\cdot x/\hbar }J^\mu (0)e^{iP\cdot x/\hbar },
\label{RJ5}
\end{equation}
we have from Eqs.(\ref{RJ2}), (\ref{RJ3}) and (\ref{RJ5}) that
\begin{equation}
J^\mu_{FI}(x)=e^{i(P_I-P_F)\cdot x/\hbar }J^\mu_{FI}(0).
\label{RJ6}
\end{equation}
yielding
\begin{equation}
-\hbar^2 (\partial_\nu \partial^\nu ) J^\mu_{FI}(x)
=(P_I-P_F)^2 J^\mu_{FI}(x).
\label{RJ7}
\end{equation}
We now have the following:
\par \noindent
{\bf Theorem 2:} {\em Real photons are radiated (or absorbed)
only by those matrix elements of
the current operator matrix \begin{math} J^\mu _{FI}(x) \end{math}
which obey the wave equation
\begin{equation}
-(\partial_\nu \partial^\nu )J^\mu _{FI\ rad}(x)=0.
\label{RJ8}
\end{equation}}
\par \noindent
{\bf Proof:} For the case of photon radiation, Eq.(\ref{RJ8}) follows from
Eqs.(\ref{RJ4}) and (\ref{RJ7}). For the case of photon absorption,
\begin{equation}
I+\gamma \to F
\label{RJ9a}
\end{equation}
the incident real photon momentum is still on the energy shell
\begin{equation}
\hbar Q=P_F-P_I\ \ \Rightarrow\ \ \hbar^2 Q^2=(P_F-P_I)^2=0.
\label{RJ9b}
\end{equation}
so that Eqs.(\ref{RJ7}) and (\ref{RJ9b}) imply Eq.(\ref{RJ8}).

An operator may be defined by its matrix elements.
Thus, the current operator may be decomposed into a
radiating and non-radiating part:
\begin{equation}
J^\mu (x)=J^\mu_{rad} (x)+J^\mu _{non-rad}(x).
\label{RJ10}
\end{equation}
The radiating part of the current obeys the wave equation
\begin{equation}
-(\partial_\nu \partial^\nu ) J^\mu_{rad}(x)=0.
\label{RJ11}
\end{equation}
Eq.(\ref{RJ11}) is the quantum electrodynamic operator generalization
of a classical wave equation constraint on classical radiating currents.
It is a central result of this work.
In Feynman diagram language, the matrix elements
\begin{math} J^\mu _{FI\ rad}(x) \end{math} appear
at a vertex in which a real photon is emitted or absorbed. The matrix
elements \begin{math} J^\mu _{FI\ non-rad}(x) \end{math} appears at each
vertex which begins or ends a virtual photon internal line.

\section{Causality}

As discussed above, a {\em virtual} photon emitted at space-time point
\begin{math} x_i \end{math} and absorbed at space-time point
\begin{math} x_f \end{math} propagates on the light cone
\begin{equation}
(x_f-x_i)^2=|{\bf r}_i-{\bf r}_f|^2-c^2(t_f-t_i)^2=0
\ \ \ {\rm (virtual\ photon)},
\label{causal1}
\end{equation}
but off the energy shell
\begin{equation}
Q^2=|{\bf k}|^2-(\omega /c)^2\ne 0
\ \ \ {\rm (virtual\ photon)}.
\label{causal2}
\end{equation}

Let us consider (as a physical example of a virtual radiation process)
the photon exchange in the M\"ossbauer effect\cite{Auriemma_88,Hans_63}.
A nucleus emits a \begin{math} \gamma \end{math}-photon which travels
a distance \begin{math} L\sim 1\  {\rm meter}\end{math} before being
absorbed by a second nucleus. The wavelength
\begin{math} \lambda =(2\pi /k) \end{math} has an uncertainty given by
\begin{math}(\Delta \lambda /\lambda )\sim (\lambda /L) \end{math}.
Typically, \begin{math}\lambda < {\rm 1\ nanometer} \end{math} so that
\begin{math}
(\Delta \lambda /\lambda )\approx (\Delta k /k) > 10^{-9}
\end{math}. On
the other hand, the frequency \begin{math} \omega  \end{math} resolution
in the M\"osbauer effect obeys
\begin{math}(\Delta \omega /\omega ) < 10^{-13} \end{math}.
The frequency \begin{math} \omega -\end{math}resolution
is more than four orders of magnitude more accurate
than is the wave number \begin{math} k-\end{math}resolution.
This can only happen if
\begin{math} \omega \ne c k  \end{math} as in Eq.(\ref{causal2}).
On the other hand, the time for the photon to
travel from one nucleus to the other is \begin{math} T=(L/c) \end{math}
in agreement with Eq.(\ref{causal1}). Thus, virtual photons appear to
be causal, with the exception of the possible backward in time processes.

For {\em real} photons, the propagation is strictly on the energy shell
\begin{equation}
Q^2=|{\bf k}|^2-(\omega /c)^2=0
\ \ \ {\rm (real\ photon)},
\label{causal3}
\end{equation}
which implies (in virtue of the uncertainty principle) propagation
off the light cone
\begin{equation}
(x_f-x_i)^2=|{\bf r}_i-{\bf r}_f|^2-c^2(t_f-t_i)^2\ne 0
\ \ \ {\rm (real\ photon)}.
\label{causal4}
\end{equation}
For example, real photons can travel space-like via the propagator
in Eq.(\ref{A9}); i.e.
\begin{equation}
{\Im m}D_{\mu \nu}({\bf r}-{\bf r}^\prime ,t-t^\prime =0)=
\frac{\eta_{\mu \nu }}{\pi |{\bf r}-{\bf r}^\prime |^2}\ .
\label{causal5}
\end{equation}
Whence comes this ability to propagate off the light cone?

\begin{figure}[htbp]
\begin{center}
\mbox{\epsfig{file=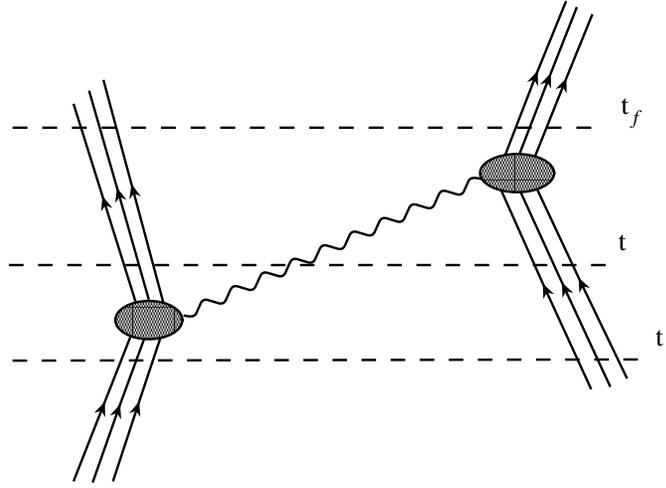,height=70mm}}
\caption{A source emits a photon shortly after an initial time $t_i$ and a
photon counter absorbs the photon shortly before a final time $t_f$. The photon
is {\em real} only for the duration of a neighborhood of the intermediate time $t$.
After time $t_f$, the photon propagation becomes a {\em virtual} exchange between
source and detector.}
\label{qedfig1}
\end{center}
\end{figure}

In his original discussion\cite{Feynman_50} concerning the relationship
between real and virtual photons, Feynman pointed out that what appears
{\em real} on a short time scale may appear {\em virtual} on a long time
scale. A photon emitted a long time ago may appear to be real in the present.
However, if at a future time the photon is absorbed then its existence
between the past and future events becomes virtual. The physical process
is shown in Fig.{\ref{qedfig1}}. When the process is completed, the momentum
conservation between source and absorber reads
\begin{eqnarray}
P_I({\rm source})+P_I({\rm detector}) &=&
P_F({\rm source})+P_F({\rm detector}), \nonumber \\
\big(P_F({\rm source})-P_I({\rm source})\big)^2 &=&
\big(P_F({\rm detector})-P_I({\rm detector})\big)^2\ne 0.
\label{causal6}
\end{eqnarray}
The photon is {\em virtual} so that \begin{math} Q^2\ne 0 \end{math} but the
propagation is on the light cone
\begin{math} (x_{source}-x_{detector})^2 =0 \end{math}.
On the other hand, during the intermediate times (in the neighborhood of
time \begin{math} t \end{math}) the photon is {\em real}. Thus
the energy shell condition \begin{math} Q^2 =0 \end{math} holds true,
but the photon velocity is not quite equal to c.

\section{Conclusions}

The {\em classical} electrodynamic notion of radiating and non-radiating
current sources has been generalized to the case of
{\em quantum} electrodynamic currents. Emission and absorption from
radiating current sources involves {\em real} photons, while
emission and absorption from non-radiating current sources involves
{\em virtual} photons. The distinction between real and virtual
can be made precise by the following conditions:
(i) {\em Real} photons propagate on the energy shell
\begin{math} Q^2=0 \end{math} but off the light cone
\begin{math} (x_f-x_i)^2\ne 0 \end{math} as required by a relativistic
uncertainty relation. (ii) Virtual photons propagate off the energy shell
\begin{math} Q^2\ne 0 \end{math} but on the light cone
\begin{math} (x_f-x_i)^2 = 0 \end{math}.

The current (in general a quantum operator) may be composed into a radiating
and non-radiating part
\begin{equation}
J^\mu (x)=J^\mu _{rad}(x)+J^\mu _{non-rad}(x).
\label{conclude1}
\end{equation}
The source for the real photons itself obeys the wave equation
\begin{equation}
-(\partial_\nu \partial^\nu )J^\mu _{rad}(x)=0,
\label{conclude2}
\end{equation}
while only the non-radiating source \begin{math} J^\mu _{non-rad} \end{math}
emits photons which propagate on the light cone.
\vskip 1.5cm
\leftline{\bf Appendix A}
\vskip .2cm
\par \noindent
Our purpose is to evaluate the photon propagator
\begin{equation}
D_{\mu \nu }(x,y) = \eta_{\mu \nu }D(x-y),
\label{A1}
\end{equation}
wherein
\begin{equation}
D(x)=\int \frac{4\pi }{Q^2-i0^+}
\ e^{iQ\cdot x}\ \frac{d^4Q}{(2\pi )^4}\ .
\label{A2}
\end{equation}
First we note that
\begin{equation}
\frac{1}{Q^2-i0^+}=i\int_0^\infty e^{-isQ^2}ds.
\label{A3}
\end{equation}
Thus
\begin{equation}
D(x)=\frac{i}{4\pi^3 }\int_0^\infty \int
\left(e^{iQx}e^{-iQ^2s}\right)d^4 Q ds.
\label{A4}
\end{equation}
Changing integration variables via the four vector
\begin{math} Q=q+(x/2s) \end{math} yields
\begin{equation}
D(x)=\frac{i}{4\pi^3 }\int_0^\infty e^{ix^2/4s}\int
\left(e^{-iq^2s}\right)d^4 q ds.
\label{A5}
\end{equation}
The Gaussian integral can be directly evaluated according to
\begin{math}
\int e^{-iq^2s}d^4 q=-i(\pi /s)^2
\end{math}, so that
\begin{equation}
D(x)=\frac{1}{4\pi }\int_0^\infty e^{ix^2/4s}
\frac{ds}{s^2}=\frac{1}{\pi }\int_0^\infty e^{iu x^2}du.
\label{A6}
\end{equation}
Finally, the propagator in space-time is given by
\begin{equation}
D_{\mu \nu}(x,y) = \eta_{\mu \nu}D(x-y)=
\frac{i\eta_{\mu \nu }}{\pi \{(x-y)^2+i0^+\}}\ .
\label{A7}
\end{equation}
Note from Eqs.(\ref{A2}), (\ref{A7}) and
\begin{equation}
\frac{1}{a^2\mp i0^+}=\frac{1}{a^2}\pm i\pi \delta(a^2),
\label{A8}
\end{equation}
follows the Fourier transform pair
\begin{eqnarray}
{\Re e}D_{\mu \nu}(x,y) &=&
\frac{1}{2}\eta_{\mu \nu }\delta\big((x-y)^2\big)
=4\pi \eta_{\mu \nu }\int \left(\frac{e^{iQ\cdot(x-y)}}{Q^2}\right)
\frac{d^4Q}{(2\pi )^4}\ ,\nonumber \\
{\Im m}D_{\mu \nu}(x,y) &=&
\frac{\eta_{\mu \nu }}{\pi \{(x-y)^2\}}
=4\pi^2 \eta_{\mu \nu }\int \delta\big(Q^2\big)e^{iQ\cdot(x-y)}
\frac{d^4Q}{(2\pi )^4}\ .
\label{A9}
\end{eqnarray}
Eqs.(\ref{A9}) implies the following: (i) On light cone propagation
\begin{math} \delta ((x-y)^2) \end{math} implies off
zero mass shell propagation \begin{math} Q^2\ne 0 \end{math} in
wave vector space. (ii) On zero mass shell propagation
\begin{math} \delta(Q^2) \end{math} implies off
light cone propagation \begin{math} (x-y)^2\ne 0 \end{math} in
space-time.

\end{document}